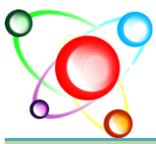

SCITECH
RESEARCH ORGANISATION

# Analytic Solution of Dirac Equation for Extended Cornell Potential Using the Nikiforov-Uvarov Method

M. Abu-Shady

Department of Applied Mathematics, Faculty of Science, Menoufia University, Egypt.

## Abstract

The extended Cornell potential which the harmonic oscillator potential is included in the original Cornell potential. The Dirac equation is solved by reducing the Dirac equation to the form of Schrodinger equation. The Nikiforov-Uvarov method is applied to obtain the energy eigenvalues and corresponding wave functions. The obtained results are important to calculate many characteristics of fermion relativistic particles.

## Introduction

The solutions of fundamental dynamical systems are interesting phenomenon in the many fields of physics. Thus, the description of atomic and subatomic physical systems using the relativistic quantum mechanics is important and impressive field. The Dirac equation plays a major role in the description of relativistic particles with spin $\frac{1}{2}$[1]. In recent years, the Dirac equation is solved for different potentials such as in Refs. [2−8]. In Ref. [2], the analytical solutions of the Dirac equation with the Cornell potential with identical scalar and vector potentials using perturbation method are investigated. In Ref. [3], the approximate solutions of the Dirac equation with scalar and vector generalized isotonic oscillators and Cornell tensor interaction using the ansatz approach are obtained. In Ref. [4], the bound energy spectrum and the corresponding generalized hypergeometric wave function of the Dirac equation for modified-Hylleraas potential under spin and pseudo spin symmetry limits in the framework of the Alhaidari-formalism are obtained. In Ref. [5], the exact solutions of Dirac equation with pseudoscalar Cornell potential using the Nikiforov-Uvarov method are obtained. In Ref. [6], the exact solution of Dirac equation for the Hartmann potential are obtained using the Nikiforov-Uvarov method. In Ref. [7], Xian-Quen et al. solved Dirac equation with new ring-shaped non-spherical harmonic oscillator when the scalar potential is equal to the vector potentail. In Ref. [8], the Dirac equation with position-dependent mass is approximately solved for the generalized Hulthen potential with any spin-orbit quantum number $k$ using the Nikiforov-Uvarov method. The Cornell potential which consists of Coulomb plus linear potentials has received a great deal of attention in particle physics. The Cornell potential was used with considerable success in models describing systems of bound states of quark and antiquark such as in Ref. [5] and references therein.

In this work, the Cornell potential is extended to include the harmonic oscillator potential which is not considered in the recent works such as in Refs. [1−8] in the framework of the Dirac equation. The Nikiforov-Uvarov method is used to calculate the energy eigenvalues and the corresponding wave functions.

The paper is organized as follows: In Sec. 2, the Nikiforov-Uvarov method is briefly explained which is used as the technique in the present paper. In Sec. 3, The energy eigenvalues and the wave functions of the Dirac equation are calculated. In Sec. 4, the summary and conclusion are presented.

## Theoretical Description of the Nikiforov-Uvarov Method

In this section, the Nikiforov-Uvarov method is briefly given, which is used as the technique to solve second-order differential equation as follows

$$\Psi^{''}(s) + \frac{\overline{\tau}(s)}{\sigma(s)}\Psi^{'}(s) + \frac{\widetilde{\sigma}(s)}{\sigma^2(s)}\Psi(s) = 0, \tag{1}$$

where $\sigma(s)$ and $\widetilde{\sigma}(s)$ are polynomials of maximum second degree and $\overline{\tau}(s)$ is a polynomial of maximum first degree. The Nikiforov-Uvarov method is given in Ref. $[9]$. The second-order differential equation which takes following form by taking the $\Psi(s)$ as follows

$$\Psi(s) = \Phi(s)\chi(s), \tag{2}$$

$$\sigma(s)\chi^{''}(s) + \tau(s)\chi^{'}(s) + \lambda\chi(s) = 0. \tag{3}$$

Eq. (1) is written as in Ref. $[10]$, where

$$\sigma(s) = \pi(s)\frac{\Phi(s)}{\Phi^{'}(s)}, \tag{4}$$

and

$$\tau(s) = \overline{\tau}(s) + 2\pi(s); \quad \tau^{'}(s) < 0, \tag{5}$$

$$\lambda = \lambda_n = -n\tau^{'}(s) - \frac{n(n-1)}{2}\sigma^{''}(s), n = 0,1,2,... \tag{6}$$

$\chi(s) = \chi_n(s)$ is a polynomial of $n$ degree which satisfies the hypergeometric equation, taking the form

$$\chi_n(s) = \frac{B_n}{\rho_n}\frac{d^n}{ds^n}(\sigma^{''}(s)\rho(s)), \tag{7}$$

where $B_n$ is a normalization constant and $\rho(s)$ is a weight function which satisfies the following equation

$$\frac{d}{ds}\omega(s) = \frac{\tau(s)}{\sigma(s)}\omega(s); \quad \omega(s) = \sigma(s)\rho(s), \tag{8}$$

$$\pi(s) = \frac{\sigma^{'}(s) - \overline{\tau}(s)}{2} \pm \sqrt{(\frac{\sigma^{'}(s) - \overline{\tau}(s)}{2})^2 - \widetilde{\sigma}(s) + K\sigma(s),} \tag{9}$$

and

$$\lambda = K + \pi^{'}(s), \tag{10}$$

the $\pi(s)$ is a polynomial of first degree. The values of $K$ in the square-root of Eq. (9) is possible to calculate if the function under the square is square of function. This is possible if its discriminate is zero.

## The Dirac Equation for the Extended Cornell Potential

The quark mass $m_\alpha$ where the index $\alpha$ refer to kind of particle ($u$, $d$, and $s$ quark in the present case) in the presence of a confining potential $V(r)$ in the Dirac equation is given $[2]$

$$\left[\alpha \cdot \mathbf{P} + \beta m_\alpha + \frac{1}{2}(1+\beta)V(r)\right]\Psi(\mathbf{r}) = E_{\alpha n}\Psi(\mathbf{r}), \tag{11}$$

where $\alpha$ and $\beta$ are the usual Dirac matrices. By decomposing the above equation in the spherical coordinates, we have the component wave function

$$\Psi(\mathbf{r}) = \begin{pmatrix} \chi(\mathbf{r}) \\ \phi(\mathbf{r}) \end{pmatrix} = \frac{1}{r} \begin{pmatrix} u(r)\Omega_k^m(\theta,\phi) \\ iv(r)\Omega_{-k}^m(\theta,\phi) \end{pmatrix}, \tag{12}$$

substituting Eq. (12) into Eq. (11) and then separating radial parts, the following radial equations are obtained

$$\frac{du(r)}{dr} = -\frac{k}{r}u(r) + \left[E_{\alpha n} + m_\alpha\right]v(r), \tag{13}$$

$$\frac{dv(r)}{dr} = \frac{k}{r}v(r) - \left[E_{\alpha n} - m_\alpha - V(r)\right]u(r), \tag{14}$$

where the $E_{\alpha n}$ is an energy eigenvalue of particle $\alpha$ and the eigenvalue of quantum number ($k$) is related to the total angular momentum quantum number $j$ by the following form

$$k = \begin{cases} -\left(j+\frac{1}{2}\right) = -(l+1) & \text{if } j=l+\frac{1}{2} \\ \left(j+\frac{1}{2}\right) = l & \text{if } j=l-\frac{1}{2} \end{cases} \tag{15}$$

Eqs. (13) and (14) are reduced to the Schrodinger equation as in $\begin{bmatrix}2\end{bmatrix}$ as follows

$$\frac{d^2}{dr^2} - \frac{k(k+1)}{r^2} + 2\varepsilon_0(\varepsilon_1 - V(r))]u(r) = 0 \tag{16}$$

where $\varepsilon_0 = \frac{1}{2}\left(E_{\alpha n} + m_\alpha\right)$ and $\varepsilon_1 = \left(E_{\alpha n} - m_\alpha\right)$. The extended Cornell potential is suggested as in Ref. $\begin{bmatrix}11\end{bmatrix}$. Thus, the potential takes the following form

$$V(r) = ar - \frac{b}{r} + cr^2, \tag{17}$$

where $a, b,$ and $c$ are arbitrary positive constants. The potential has distinctive features of strongly interaction, namely, the asymptotic freedom and the confinement are represented in the first and the second terms, respectively. The combined two terms are called the Cornell potential. The third term $cr^2$ is called the harmonic oscillator potential which defines the particle with mass $m$ oscillates with frequency $\omega$ where the parameter $c$ is proportional with $m$ and $\omega^2$. The harmonic oscillator potential plays important role on the effect of the quarkonium properties as in Refs. $\begin{bmatrix}11,12\end{bmatrix}$ By substituting Eq. (17) into Eq. (16), we obtain

$$\frac{d^2}{dr^2} + 2\varepsilon_0(\varepsilon_1 - ar + \frac{b}{r} - cr^2 - \frac{k(k+1)}{2\varepsilon_0 r^2})]u(r) = 0. \tag{18}$$

Let assume $r = \frac{1}{x}$ and $r_0$ a characteristic radius of the meson. Then the scheme is based on the expansion of $\frac{1}{x}$ in a power series around $r_0$, i. e. around $\delta = \frac{1}{r_0}$ in $x$ space (for details, see Ref. [10] and the references therein). Eq. (18) takes the following form

$$\frac{d^2}{dx^2} + \frac{2x}{x^2}\frac{d}{dx} + \frac{2\varepsilon_0}{x^4}(-A + Bx - C_1 x^2)]u(x) = 0, \tag{19}$$

where, $A = -\varepsilon_0(\varepsilon_1 - \frac{3a}{\delta} - \frac{6c}{\delta^2})$, $B = \varepsilon_0(\frac{3a}{\delta^2} + \frac{8c}{\delta^3} + b)$, and $C_1 = \varepsilon_0(\frac{a}{\delta^3} + \frac{c}{\delta^4} + \frac{k(k+1)}{2\varepsilon_0})$. By comparing Eq. (19) and Eq. (1), we find $\bar{\tau}(s) = 2x$, $\sigma(s) = x^2$, and $\widetilde{\sigma}(s) = 2\varepsilon_0(-A + Bx - C_1 x^2)$. Hence, the Eq. (19) satisfies the conditions in Eq. (1). By following technique is mentioned in Sec. 2 as well as in Ref. $\begin{bmatrix} 10 \end{bmatrix}$, the energy eigenvalue of Eq. (18) is given

$$\varepsilon_1 = \frac{3a}{\delta} + \frac{6c}{\delta^2} - \frac{2\varepsilon_0(\frac{3a}{\delta^2} + b + \frac{8c}{\delta^3})^2}{[(2n+1) \pm \sqrt{1 + \frac{8\varepsilon_0 a}{\delta^3} + 4k(k+1) + \frac{24\mu c}{\delta^4}}]^2}. \tag{20}$$

In Ref. $\begin{bmatrix} 10 \end{bmatrix}$, the Cornell potential is only used to obtain the energy eigenvalues. Hence, the energy eigenvalues in Ref. $\begin{bmatrix} 10 \end{bmatrix}$ is a particular case from Eq. (20) when the parameter $c = 0$. By following the steps in the section 2. The radial of wave function of Eq. (18) takes the following form

$$u_{nk}(r) = C_{nk} \, r^{-\frac{B}{\sqrt{2A}} - 1} e^{\sqrt{2A}r} (-r^2 \frac{d}{dr})^n (r^{-2n + \frac{2B}{\sqrt{2A}}} e^{-2\sqrt{2A}r}), \tag{21}$$

where $C_{nk}$ is a normalization constant that is determined by $\int |R_{nk}(r)|^2 \, d\mathbf{r} = 1$ and

$$A = -\varepsilon_0(\varepsilon_1 - \frac{3a}{\delta} - \frac{6c}{\delta^2}), \qquad B = \varepsilon_0(\frac{3a}{\delta^2} + b + \frac{8c}{\delta^3}). \tag{22}$$

## Summary and Conclusion

In this work, the modified Cornell potential is suggested, which plays an important role for describing the interaction between quark and antiquark at short distances in the bound state of meson as in Refs. $\begin{bmatrix} 11,12 \end{bmatrix}$. The Dirac equation is solved for the modified Cornell potential, where the Dirac equation is reduced to the Schrodinger equation as in Ref. $\begin{bmatrix} 2 \end{bmatrix}$. The Nikiforov-Uvarov method is applied to obtain the energy eigenvalues and corresponding wave functions. The advantage of the present work that the modified Cornell potential is not considered in the many recent works that mentioned in the section (1) by using the Nikiforov-Uvarov method.

We conclude that the obtained results are important to calculate many characteristics of fermion relativistic particles.

## References

(1) A. D. Alhaidari, H. Bahlouli, and A. Al-Hasan, Phys. Lett. A **349**, 87 (2006).

(2) L. A. Trevisan, Carlos Mirez, and F. M. Andrade, Few-body Syst. **55** 1055 (2014).

(3) H. Hassanabadi, E. Maghsoodi, Akpan N. Ikot, and S. Zarrinkamar, Adv. High Energ. Phys. **2014**, 831938 (2014).

(4) A. N. Ikot, E. Maghsoodi, O. A. Awoga, S. Zarrinkamar, H. Hassanabadi, Quant. Phys. Lett. **3**, 7 (2014).

(5) M. Hamzavi and A. A. Rajabi, Chinese Phys. C **37**, 103102 (2013).

(6) M. Hamzavi, H. Hassanabadi, and A. A. Rajabi, Inter. J. Mod. Phys. E, 19, 2189 (2010).

(7) H. Xian-Quen, L. Guoug, W. Yhi-Min, N. Lion-Bin, and M. Yan. Commun. Theor. Phys. **53**, 242 (2010).

(8) A. Arda, R. Sever, and C. Tezcan, Cent. Eur. J. Phys. **8**, 843 (2010).

(9) Af. Nikiforov Uvarov, "Special Functions of Mathematical Physics" Birkhauser, Basel (1988).

(10) S. M. Kuchin and N. V. Maksimenko, Univ. J. Phys. Appl. **7**, 295 (2013).

(11) R. Kumar and F. Chand, Commun. Theor. Phys. **59**, 528 (2013).

(12) P. Gupta and I. Mechrotra, J. Mod. Phys. **3**, 1530 (2012).